\documentclass[twocolumn,aps,nopacs]{revtex4}

\usepackage{amsmath}
\usepackage{graphicx}
\usepackage{dsfont}

\begin{document}

\title{Conditional cooling limit for a quantum channel going through an incoherent environment}

\author{Ivo Straka\textsuperscript{\textdagger}}
\email{straka@optics.upol.cz}

\author{Martina Mikov\' a}
\altaffiliation{These authors contributed equally to this work.}

\author{Michal Mi\v cuda}
\author{Miloslav Du\v sek}
\author{Miroslav Je\v zek}
\author{Radim Filip}

\affiliation{Department of Optics, Palack\' y University, 17. listopadu 1192/12, 771~46 Olomouc, Czech Republic}

\begin{abstract}
We propose and experimentally verify a cooling limit for a quantum channel going through an incoherent environment. The environment consists of a large number of independent non-interacting and non-interfering elementary quantum systems -- qubits. The qubits travelling through the channel can only be randomly replaced by environmental qubits. We investigate a conditional cooling limit that exploits an additional probing output. The limit specifies when the single-qubit channel is quantum, i.e. it preserves entanglement. It is a fundamental condition for entanglement-based quantum technology.
\end{abstract}

\maketitle

\thispagestyle{empty}

\section*{Introduction}

Entanglement of two quantum bits (qubits) is a key feature to understand the microscopic world \cite{ent1,ent2,ent3,ent4} and also a basic resource of an important branch of quantum technology \cite{ent5}. Entanglement is however a fragile resource. It is sensitive to a coupling of an entangled qubit to unavoidably present surrounding environment \cite{Zurek}. The environment represents a class of quantum channels \cite{NS}, which may reduce the entanglement or possibly even break it completely. In our analysis, we regard such channel as quantum if it is not entanglement breaking \cite{Ruskai}.

We consider the most common incoherent environment containing multiple non-interacting and non-interfering qubits. In such a case, there is only a single mechanism reducing the entanglement. The qubit going through the environment can be lost and another fully incoherent qubit with noisy features can be found at the channel output. The incoherent qubits in the environment are considered as distinguishable from the qubits being transferred \cite{indis}, but technically indistinguishable. It was proven that an incoherent environment containing only a single qubit is already sufficient to break entanglement in the channel \cite{inc1,inc2} and also substantially reduce the direct applicability in quantum technology \cite{inc3}. Remarkably, if the entanglement remains after the channel, a conditional entanglement distillation can be used to approach the maximal entanglement and recover the quantumness of the channel \cite{dist1,dist2}. However, if the entanglement is completely lost, no entanglement distillation can help to fully recover the ideal quantum channel. The channel becomes entanglement breaking \cite{break} and the environment degrades the channel to classical one. The preservation of entanglement is therefore a fundamental limit for all the quantum channels in entanglement-based technology.

To test this limit for the channel, we employ a maximum entanglement between a reference qubit R and the qubit entering the channel. At the output of the channel, we get qubit A leaving and verify the entanglement between qubits R and A. We consider the environment to be a large reservoir of qubits. As a result, we can represent the state of each environmental qubit by a mixed state $\mathcal{E}$. The diagonal basis consists of a ground state $|\psi\rangle$ and an excited state $|\psi_\perp\rangle$, which is populated with a probability $p_T$. For an environment at thermal equilibrium, the basis states become energy eigenstates with $p_{T}\propto \exp\left(-\frac{\Delta E}{k_BT}\right)$, where $\Delta E$ is a difference in energy between the ground and excited states of the qubit, $T$ is the temperature of the environment and $k_B$ is the Boltzmann constant. 
First, a straightforward step is to find the maximum $p_T$, or the maximum temperature $T$ of the environment, which still guarantees non-vanishing entanglement after the channel. This deterministic approach yields an unconditional limit on a cooling of the environment to keep the channel quantum.
In the previous works \cite{inc1,inc2,inc3}, only single incoherent non-interfering and non-interacting qubit in the environment was analysed and the limit was experimentally tested. However, a realistic environment is typically complex, consisting of many non-interacting and non-interfering noisy qubits. Therefore, this case is the main subject of our analysis here.

\begin{figure}
\centerline{\includegraphics[width=\linewidth]{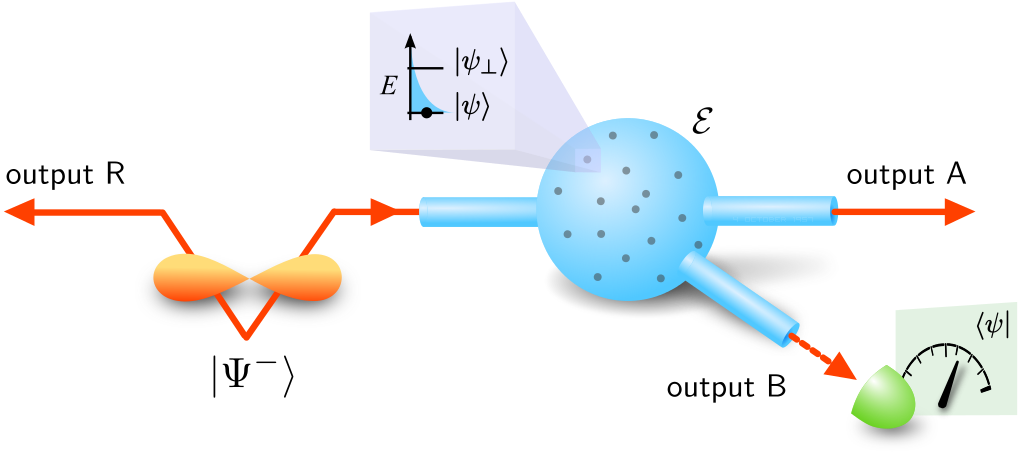}}

\caption{The schematics of the incoherent channel. Since we assume one qubit in each output, we will distinguish the qubits and outputs commonly by R, A, B. A qubit entangled with qubit R is propagated through an incoherent environment. The environment consists of many qubits in a mixture of the states $|\psi\rangle,|\psi_\perp\rangle$. A projective measurement is carried out on output B, conditioning an entanglement preservation on output A.}

\label{scheme}
\end{figure}

In this paper, we derive and experimentally verify a conditional cooling limit for quantum channel with incoherent environment with many independent noisy qubits. 
To derive the conditional limit, we extract an auxiliary qubit B from the incoherent environment, as is depicted in Fig.~1. It can be advantageously used to herald more entanglement between qubits R and A, thus beating the unconditional cooling limit. In this approach, the entanglement is fully broken only if the qubit entering the environment is lost and two completely incoherent environmental qubits  appear at the outputs of the environment. The probability of the qubit being lost is $P_L$. We derive a limit on a joint error represented by the product $p_TP_L$, depending on temperature $T$ for thermal environment. The limit is given by the probability $P_S$ of successful implementation of the ideal quantum channel. We experimentally verify this fundamental cooling limit using a quantum optics experiment with a simulated, controllable environment. However, it must be noted that our analysis is not limited solely to thermal environments or single photons; it is applicable to any environment consisting of independent non-interacting and non-interfering qubits. The conditional cooling limit for qubit quantum channel is universal, widely applicable for different physical platforms.

\section*{Unconditional cooling limit}

To derive an unconditional cooling limit, we ignore the possibility of access by any auxiliary output in the environment. We focus only on the case when a single qubit is present in the output A. An environmental qubit is  considered in the state $\mathcal{E}=(1-p_T)|\psi\rangle\langle\psi|+p_T|\psi_\perp\rangle\langle\psi_\perp|$, where $|\psi\rangle$ is the ground state of the environment and $p_T \leq \frac{1}{2}$ is the probability of a thermal excitation of the environment. To derive the maximum of the probability $p_T$ (for thermal environment, a maximum of temperature $T$), we consider a maximally entangled state $|\Psi^-\rangle=\frac{1}{\sqrt{2}}\left(|\psi\rangle|\psi_\perp\rangle-|\psi_\perp\rangle|\psi\rangle\right)$ between the reference qubit and the channel qubit, written in the basis of the environmental state $\mathcal{E}$. This state was chosen as a probe for the channel, because any other state would yield a more or equally strict condition on $p_T$. As a result, the condition for $p_T$ represents a necessary criterion for the channel itself to be quantum, preserving entanglement.

After the channel, the state $|\Psi^-\rangle\langle\Psi^-|$ changes to the unconditional state
\begin{equation}\label{rho.u}
P_S|\Psi^-\rangle \langle\Psi^-|_\text{R,A}+(1-P_S)\left(\frac{1}{2}\mathds{1}_\text{R}\otimes\mathcal{E}_\text{A}\right),
\end{equation}
where $P_S$ is the probability of successful transmission of the maximally entangled state. The entanglement is preserved for any $p_T$ only if $P_S>1/3$. Otherwise, the general implicit condition on a quantum channel is
\begin{equation}\label{cond1}
\frac{\sqrt{p_T(1-p_T)}}{1+\sqrt{p_T(1-p_T)}}<P_S,
\end{equation}
which follows from \eqref{rho.u} using PPT entanglement criteria \cite{ent4}. When conventional cooling techniques are used to reduce the temperature of the environment, the limit of small residual $p_T \ll 1$ becomes available. The inequality \eqref{cond1} can be approximated by
\begin{equation}\label{appur1}
\frac{p_T}{P_S^2}<1 \quad \text{for} \quad p_T\ll 1,
\end{equation}
which is the very simple condition determining how well the environment has to be cooled down to still allow a quantum channel. For $P_S \ll 1$, the approximation \eqref{appur1} is also valid. If the condition \eqref{appur1} is satisfied, the condition \eqref{cond1} is as well. The inequality \eqref{appur1} is reminiscent of a condition widely used to verify nonclassicality of single-photon sources \cite{Mandel,Grangier,Filip}. Here, $P_S$ stands for the success probability of transmitting a pure singlet $|\Psi^-\rangle\langle\Psi^-|$ through the environment, and $p_T$ is the measure of error, represented by a random thermal excitation of an environmental qubit.

\section*{Conditional cooling limit}

Suppose now that an auxiliary qubit B, extracted from the environment, can be detected to herald the state in the outputs R and A. No coherent operation between the qubits B and R--A will work since B needs not be coherent with R and A. A measurement on B is therefore the only way how to improve the unconditional limit \eqref{cond1}. Differently to the previous case, a random mixture of three elementary effects is present now in the environment. With a success probability $P_S$, the channel yields the qubit entangled with R to the output A, so the singlet is transferred unchanged. This leaves a noisy state $\mathcal{E}$ for the auxiliary output B. With a flip probability $P_F$, the qubit entangled with R emerges from the auxiliary output B, while on A we find the noisy state $\mathcal{E}$. With a probability of loss $P_L$, two completely incoherent qubits $\mathcal{E} \otimes \mathcal{E}$ appear at the outputs A and B. This last process is extremely destructive, the entanglement with reference R is lost in the environment. All three effects are considered to be technically indistinguishable from outside of the environment.

Due to $P_S+P_F+P_L=1$, we can use only $P_S$ and $P_L$ to fully characterize the resulting state

\begin{eqnarray}\label{3state}
\rho_\text{R,A,B}&=& P_S|\Psi^-\rangle\langle\Psi^-|_\text{R,A}\otimes \mathcal{E}_\text{B} + \\\nonumber
& & P_F |\Psi^-\rangle\langle\Psi^-|_\text{R,B}\otimes \mathcal{E}_\text{A} + P_L \frac{1}{2}\mathds{1}_\text{R} \otimes \mathcal{E}_\text{A} \otimes \mathcal{E}_\text{B}
\end{eqnarray}
describing a broad class of physical situations at many experimental platforms.

To detect the state of qubit B, we assume general projective measurement $|\Phi\rangle\langle \Phi|_\text{B}$. For any $p_T<\frac{1}{2}$, the optimal strategy is to implement the projector $|\Phi\rangle\langle\Phi|_\text{B}=|\psi\rangle\langle\psi|_\text{B}$ on the more probable ground state.

The resulting state is then
\begin{eqnarray}\label{localized state}
\frac{1}{N}\left[(1-p_T)\left(P_S|\Psi^-\rangle\langle\Psi^-|_\text{R,A} + P_L \frac{1}{2} \mathds{1}_\text{R} \otimes \mathcal{E}_\text{A}\right) + \right. \nonumber\\
\left. \frac{1}{2} P_F |\psi_\perp\rangle\langle\psi_\perp|_\text{R} \otimes \mathcal{E}_\text{A}\right],
\end{eqnarray}
where $N=(1-p_T)(1-P_F)+P_F/2$.

In this case, the conditional state preserves entanglement between R and A if the probability $P_S$ of success satisfies
\begin{equation}\label{cond2}
P_S>\frac{1}{2}\left(\sqrt{P_{TL}(4-3P_{TL})}-P_{TL}\right).
\end{equation}
The joint error magnitude $P_{TL}=p_TP_L$ simply incorporates both undesirable sources of probabilistic error: the thermal excitation of the environment $p_T$ and the probability of entanglement loss $P_L$. Without either of these errors, the channel would always be quantum. The case of $P_L=0$ was predicted and already experimentally tested in Refs.~\cite{inc1,inc2}. However, it did not test an environment with more than a single qubit. In this case of $P_L>0$, the projection on qubit B allows us to compensate $P_L$ by cooling the environment in order to fulfil condition \eqref{cond2}.

At a high temperature limit $p_T\approx \frac{1}{2}$ and for small $P_L\ll 1$, the condition $P_S>\sqrt{P_L/2}$
substitutes the more strict condition $P_S>\frac{1}{3}$ for a guaranteed entanglement-preserving channel.
More interestingly, for very cold environments with $p_T\ll 1$, the condition \eqref{cond2} can be well approximated by
the condition
\begin{equation}\label{appur2}
\frac{P_{TL}}{P_S^2}<1 \quad \text{for} \quad p_T\ll 1,
\end{equation}
simply comparable to \eqref{appur1}. The probability $p_T$ in \eqref{appur1} is now substituted by $P_{TL}=p_TP_L$, which becomes lower when the environment is cooled down. The condition \eqref{appur2} represents the conditional cooling limit for quantum channel through incoherent many-qubit environment.

\section*{Photonic simulation}

Experimentally, we have no way of directly measuring the channel parameters $P_S$ and $P_L$, because we have no access to the noise process. We are, however, able to observe the preservation of entanglement. To positively recognize the limits of the quantum channel, we need the necessary conditions \eqref{cond1}, \eqref{cond2} to be sufficient as well. Therefore, we use the maximally entangled singlet state as a probe, like we did in the theoretical analysis.

\begin{figure}
\centerline{\includegraphics[width=\linewidth]{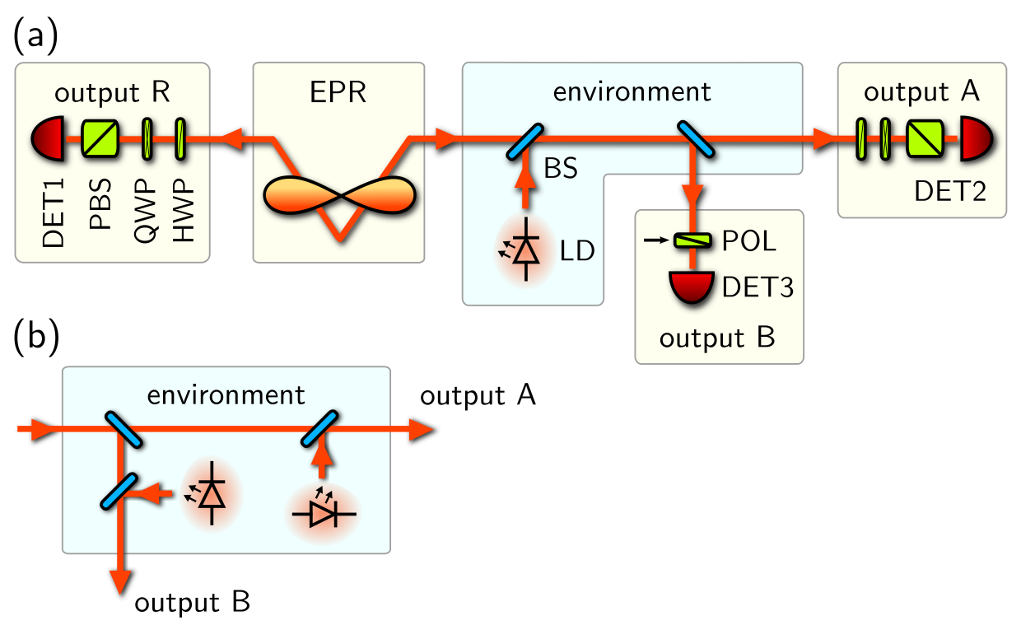}}
\caption{a) The schematic of the experiment. b) An extension allowing more general setting of the parameters $P_S, P_L, p_T$.}
\label{setup}
\end{figure}

In previous work \cite{inc1}, only single-photon noise was considered. This case is represented in our parametric space by the plane $P_L=0$. Our proposed simulator covers a more general case of non-zero $P_L$. For our proof-of-principle measurement, we used the setup shown on Fig. \ref{setup}a. The simulated parameters are then bound by $2P_S + P_L = 1$. If one needs to simulate a general set of $P_S, P_F, P_L$, one would simply use the environment shown on Fig. \ref{setup}b.

\section*{Experimental simulation}

As mentioned, the bounds can be tested in any basis. Our experiment employed the polarization basis $|H\rangle, |V\rangle$. A two-photon singlet polarization state $ |\Psi^-\rangle = \frac{1}{\sqrt 2}(|HV\rangle-|VH\rangle)$ was conditionally generated by a collinear type-II parametric down-conversion in a BBO crystal. One photon was then propagated through a noisy channel, where a polarization state $\mathcal{E}_{|H\rangle,|V\rangle}$ was incoherently coupled as noise on a 50:50 beam splitter. An attenuated laser diode was employed as the source of this independent noise. Probing was done by splitting the signal on the second 50:50 beam splitter and detecting an auxiliary photon on detector DET3. Eventually, both output ports R, A were subjected to a polarization projection. Silicon avalanche photo-diodes (Excelitas) were used for detection. For data acquisition, a time-tagging module (qutools) was used.

We carried out a state tomography to reconstruct the state on the outputs R, A. Every measurement was conditioned by all three detectors clicking. The coincidences between R and A filter out residual single photons present in each arm, that are inherent to every photon-pair source that includes optical loss. The detection on the auxiliary output B represents a successful extraction of a particle from the noisy environment. The polariser POL serves as a projective measurement $|H\rangle\langle H|_\text{B}$.

\begin{figure}
\centerline{\includegraphics[width=\linewidth]{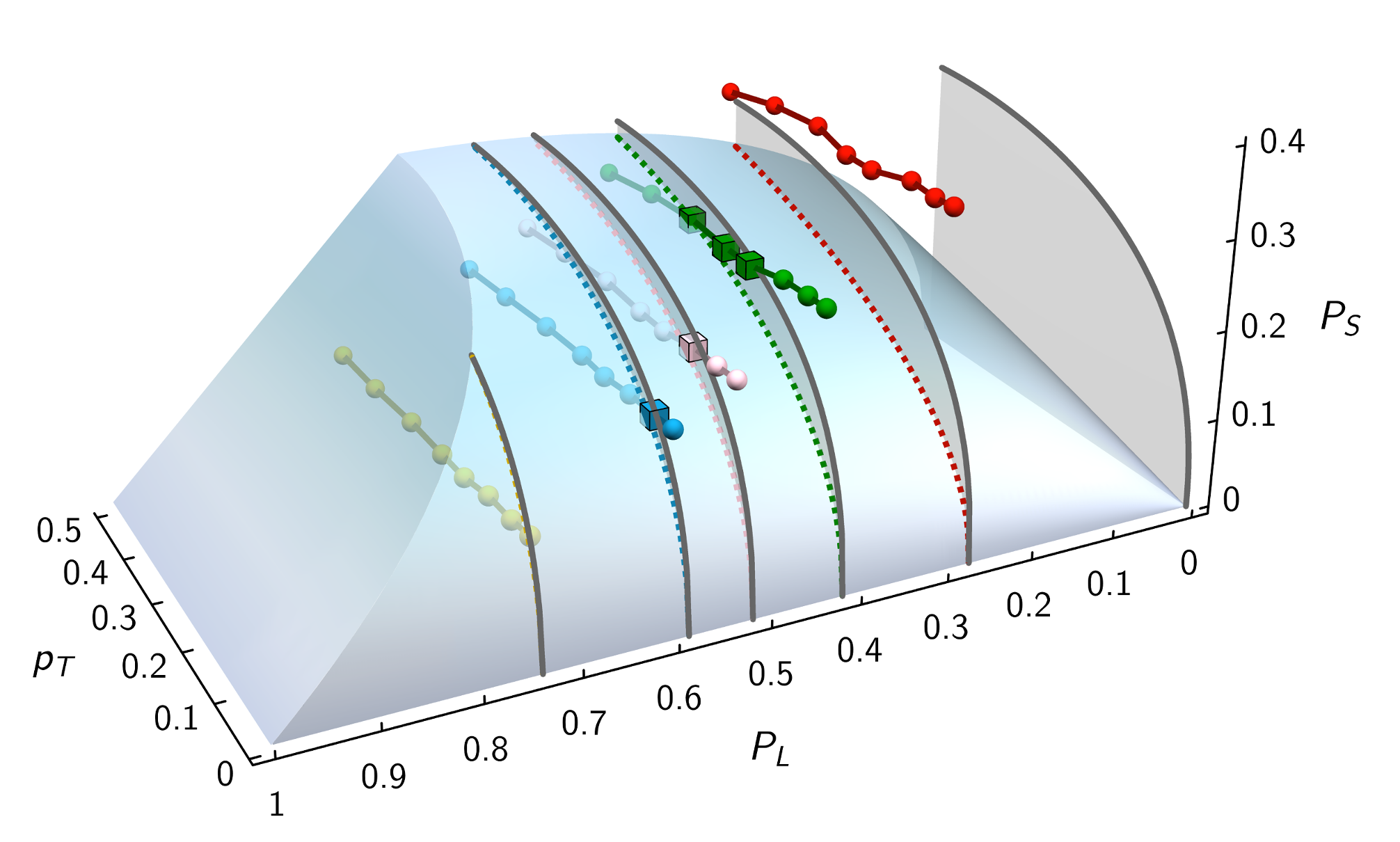}}
\caption{Measured parameters $p_T, P_S, P_L$ conditioned by a detection on output B. Each dataset represents various temperatures of the noise for a certain noise intensity, which determines $P_S$ and $P_L$. Solid grey lines represent the unconditional limit \eqref{cond1} and coloured dashed lines represent the conditional limit \eqref{cond2} for the respective $P_L$ values. The space is limited by the condition $P_L + P_S \leq 1$, which visibly cuts both separability limits for higher $P_L$. For a clear visualisation of numerical values, see Fig. \ref{fig2D}.}
\label{fig3D}
\end{figure}

\begin{figure}
\centerline{\includegraphics[width=\linewidth]{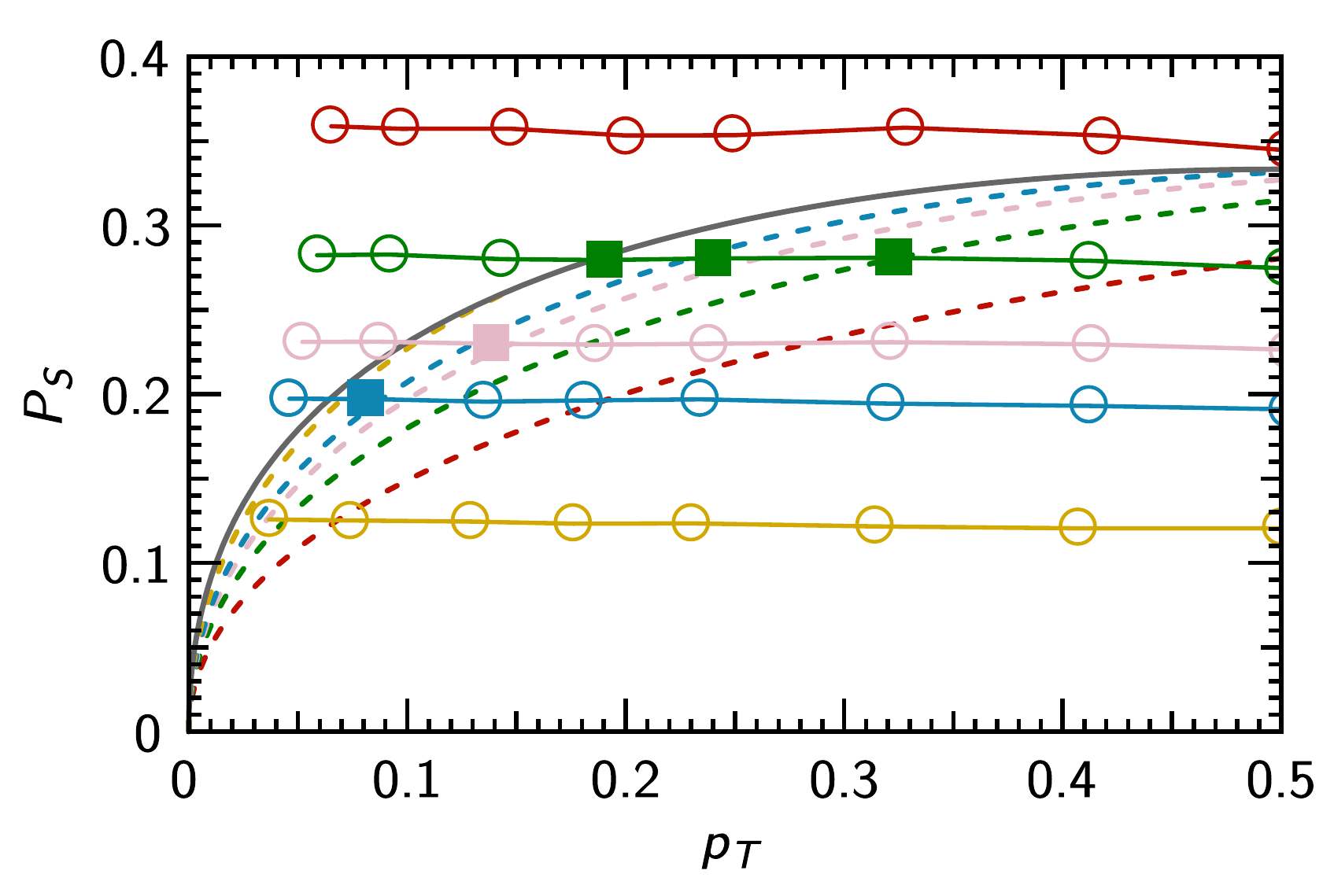}}
\caption{The data from Fig. \ref{fig3D}, where each colour denotes a certain value of $P_L$. Square points represent the states, for which the entanglement was preserved only conditionally.}
\label{fig2D}
\end{figure}

\section*{Generated quantum state}

The output state generated by the simulation is approximately of the form \eqref{localized state}. In terms of experimental parameters, $P_S = \frac{1}{2+\mathcal{R}}, P_L = \frac{\mathcal{R}}{2+\mathcal{R}}$, where $\mathcal{R} = R_N R_S\tau/R_{\Psi^-}$. $R_{\Psi^-}$ is the count rate of the singlet state,  $R_S$ the rate of single photons in output R and $R_N$ the rate of noise. $\tau$ is the coincidence window.

In order to manipulate the two parameters $p_T$ and $\mathcal{R}$ effectively, we performed a measurement for $\mathcal{E}=|H\rangle\langle H|,|V\rangle\langle V|$, separately. By probabilistically mixing the detections, we have control over $p_T$ without the need to carry out a full state tomography for each value. On the side of $\mathcal{R}$, the most convenient parameter to change is the coincidence window $\tau$.

In Figures \ref{fig3D} and \ref{fig2D}, the measured data are shown. Fig. \ref{fig3D} best illustrates the parametric space spanned by $P_S,P_L,p_T$. The blue surface represents the conditional bound for separability \eqref{cond2}, while the thick grey lines belong to the surface representing the unconditional bound \eqref{cond1}. Therefore, spherical data points below the blue surface represent quantum states, where the entanglement is lost. Spherical points above the surface are the states, which remain entangled unconditionally. Cube points are the states between the two conditions---separable unconditionally, but entangled using the auxiliary projection.

Each line represents a certain kind of noisy channel parametrised by $P_S, P_L$, experimentally set using the coincidence window $\tau$. The points along the data lines correspond to various purity of the noise state $\mathcal{E}$, quantified by $p_T$. The data illustrate that for sufficiently low temperature of the noise, here represented by the purity of $\mathcal{E}$, the noisy channel is not entanglement-breaking.

\section*{Accessible channel parameters}

In our demonstration, we used only two degrees of freedom, $p_T$ and $\mathcal{R}$, because we did not need to simulate any specific $P_S, P_L$. To obtain the third degree of freedom, one needs to couple the incoherent noise to outputs A and B separately, with different intensities (see Fig. \ref{setup}b). Assuming an ideal EPR source, this allows covering the whole parametric space.

For a realistic entanglement source, however, the regions of $P_{S,F,L} \rightarrow 0$ must be discussed. The first limit is the lower bound on the overall coupled noise intensity, which needs to be much stronger than single photons generated by the EPR source. This approximation is necessary to experimentally obtain the state \eqref{localized state}. As a result, one can reach the region of $P_L \rightarrow 0$ either by reducing inherent losses in the EPR source, or by using nonclassical single-photon noise \cite{inc1}.

The opposite limit of a strong noise gives the bounds $P_L < P_S/r_{\Psi^-}$ and $P_L < (1-P_S)/(1-r_{\Psi^-})$, where $r_{\Psi^-} = R_{\Psi^-}/(4 R_S)$ is the singlet generation rate relative to single-photon background in output R. In this case, if either $P_S,P_F \rightarrow 0$, one needs only to sufficiently attenuate the signal before the environment to decrease the ratio $r_{\Psi^-}$.

\section*{Summary}

We presented necessary conditions to preserve entanglement propagated through an incoherent environment. We showed that entanglement can be preserved using a local auxiliary projection even for a many-particle environment. For sufficient cooling of a thermal environment, these conditions are reduced to simple error-success ratios \eqref{appur1}, \eqref{appur2}. We also presented a photonic experiment as a convenient way to simulate this scenario for an arbitrary set of parameters. Finally, we experimentally verified that entanglement can be conditionally preserved using the proposed probing technique.

The conditional cooling limit is also an interesting and relevant problem for other quantum channels coupled to a thermal environment. For a single-mode continuous-variable channels with thermal noise, a conditional correction restoring the entanglement-preserving nature of the channels was already proposed and experimentally tested in an all-optical simulator \cite{envc}. Recently, the conditional approach to ground state preparation was experimentally used to cool down a mechanical cantilever in a pulsed regime \cite{mech1,mech2}. The conditional preparation of entanglement between two mechanical systems was proposed in \cite{ment1,ment2,ment3,ment4,ment5} and is now being investigated experimentally. In future work, it would be therefore stimulating to extend the conditional cooling limit to quantum optomechanical systems.

\section*{Acknowledgements}

This work was supported by the Czech Science Foundation (13-20319S). M. Mikov\'a acknowledges the support of Palack\'y University (IGA-PrF-2015-005).

\section*{Author contributions}

R. F. provided the theoretical concept and analysis. M. Mikov\' a, M. J., I. S., and M. Mi\v cuda performed the experiment. M. J. and M. D. supervised and coordinated the experiment. I. S. and R. F. wrote the main text. All authors were involved in editing and revising the manuscript.

\end{document}